\documentclass[twocolumn]{jpsj3}

\usepackage{color}
\usepackage{times}
\usepackage{graphicx}

\title{Impurity Effects in Nodal Extended $\mib{s}$- and Nodeless $\mib{d}$-Wave Superconductors:\\
Gap Symmetry of BiS$_2$-Based Layered Superconductors}

\author{Akihiro Ichikawa and Takashi Hotta}

\inst{Department of Physics, Tokyo Metropolitan University,
Hachioji, Tokyo 192-0397, Japan}

\recdate{\today}

\abst{
In BiS$_2$-based layered superconductors,
the existence of gap nodes on Fermi-surface curves has been
suggested from angle-resolved photoemission spectroscopy measurements,
whereas the conventional $s$-wave gap has been proposed from
measurements of superfluid density and thermal conductivity.
To reconcile these two distinct experimental results of the gap node,
we investigate nonmagnetic impurity effects in the superconductor
with a disconnected pocketlike Fermi-surface structure.
Then, we claim that the seemingly contradictory situation concerning the gap node
is resolved by a concept of dirty nodal extended $s$-wave superconductivity.
Provided that it is unnecessary to consider the nodes of the gap, at first glance,
the conventional $s$-wave gap seems to be a unique solution,
but in the pocketlike Fermi-surface topology,
a nontrivial possibility of a nodeless $d$-wave superconductor is pointed out.
To clarify the gap symmetry, we propose to perform experiments
on nuclear magnetic relaxation rate $T_{1}^{-1}$ in BiS$_2$-based layered superconductors.
}

\begin{document}
\maketitle

\section{Introduction}

In 2012, Mizuguchi {\it et al.} discovered the new BiS$_2$-based layered superconductor
LaO$_{1-x}$F$_x$BiS$_2$.\cite{Mizuguchi1,Mizuguchi2,Mizuguchi3}
The mother compound LaOBiS$_2$ is insulating, while owing to the substitution of F for O,
electrons are doped into the BiS$_2$ layer.
Then, the system becomes metallic and superconductivity occurs at low temperatures.
It has been experimentally known that the superconducting transition temperature
$T_{\rm c}$ is maximum at $x=0.5$.\cite{Mizuguchi3}.
In the sample synthesized under high pressures,
the highest $T_{\rm c}$ among the BiS$_2$ family has been obtained.\cite{Mizuguchi4}
The onset $T_{\rm c}$ is 11.1 K and the temperature at which the resistivity
becomes zero is 8.5 K.

Concerning the mechanism of superconductivity in new materials,
an important clue has been frequently obtained from the information
on the gap symmetry revealed by the measurement of physical quantities
and theoretical research on the gap structure.
In the case of BiS$_2$-based layered superconductors,
we notice that the results are consistent with $s$- or extended $s$-wave pairing
suggested from the temperature dependence in the superfluid densities
of Bi$_4$O$_4$S$_3$,\cite{Srivastava}
LaO$_{0.5}$F$_{0.5}$BiS$_2$,\cite{Lamura}
and NdO$_{1-x}$F$_{x}$BiS$_2$.\cite{Jiao}

Among them, here, we focus on NdO$_{1-x}$F$_{x}$BiS$_2$.
As mentioned above, it has been reported that the gap symmetry is
consistent with $s$- or extended $s$-wave for both $x=0.3$ and $0.5$
in the measurement of superfluid density.\cite{Jiao}
Also, from the thermal transport measurements,
a conventional $s$-wave gap has been considered to be realized
in NdO$_{0.71}$F$_{0.29}$BiS$_2$.\cite{Yamashita}
On the other hand, a different result of the gap structure has been
reported for NdO$_{0.71}$F$_{0.29}$BiS$_2$.
Namely, the existence of the gap node on the Fermi-surface curves
has been suggested from angle-resolved photoemission spectroscopy (ARPES)
measurements.\cite{Ota}
At first glance, these results seem to contradict each other.
It is an important and challenging task to find a way to reconcile those results
of the symmetry of the gap function from a theoretical viewpoint.

In this paper, we investigate nonmagnetic impurity effects on
BiS$_2$-based layered superconductors
by evaluating the density of states (DOS),
nuclear magnetic relaxation rate $T_1^{-1}$,
and the superfluid density $\rho_{\rm s}$
for both nodal extended $s$-wave and nodeless $d$-wave gap functions
within a self-consistent $T$-matrix approximation.
On the basis of a concept of dirty nodal extended $s$-wave superconductors,
it is claimed that the existence of the node of the gap
on the Fermi-surface curve does not contradict
the $s$-wave-like temperature dependence of $\rho_{\rm s}$.
Provided that the gap nodes on the Fermi-surface curves are ignored,
the nodeless $d$-wave superconductor
becomes another candidate in the present Fermi-surface topology,
in addition to the conventional $s$-wave gap.

The paper is organized as follows.
In Sect.~2, we show our model and the formulation for the
impurity effects in the superconducting state.
We also explain the $T_{\rm c}$ reduction due to nonmagnetic impurities
and the Fermi-surface structure of BiS$_2$-based layered superconductors.
In Sect.~3, we show our calculation results of
the physical quantities for both extended $s$-
and nodeless $d$- wave superconductors.
Finally, in Sect.~4, we summarize this paper
and discuss the possible relationship of our scenario on dirty superconductors
with the experimental results.
We also propose the experimental measurement of $T_{1}^{-1}$
in BiS$_2$-based layered superconductors.
Throughout this paper, we use such units as $\hbar=k_{\rm B}=1$.

\section{Model and Formulation}

\subsection{Model Hamiltonian}

In this paper, we consider the Hamiltonian
\begin{equation}
H=H_0+H_{\rm imp},
\end{equation}
where $H_0$ denotes the effective Hamiltonian without impurities,
whereas  $H_{\rm imp}$ indicates the non-magnetic impurity term.
The first term $H_0$ is given by
\begin{equation}
H_0=\sum_{\mib{k}\sigma}\varepsilon_{\mib{k}}
c^{\dag}_{\mib{k}\sigma}c_{\mib{k}\sigma}
+\sum_{\mib{k},\mib{k'}}V_{\mib{k},\mib{k'}}
c^{\dag}_{\mib{k}\uparrow}c^{\dag}_{\mib{-k}\downarrow}
c_{\mib{-k'}\downarrow}c_{\mib{k'}\uparrow},
\end{equation}
where $\varepsilon_{\mib{k}}$ is the kinetic energy of electrons
with the wave vector $\mib{k}$, 
$c_{\mib{k}\sigma}$ is the annihilation operator of electrons
with $\mib{k}$ and spin $\sigma$,
and $V_{\mib{k},\mib{k'}}$ is the effective attraction between electrons.

The second term $H_{\rm imp}$ is given by
\begin{equation}
H_{\rm imp}=U \sum_{i,\sigma}
c^{\dag}_{\mib{r}_i \sigma}c_{\mib{r}_i \sigma},
\end{equation}
where $U$ denotes the potential due to nonmagnetic impurities,
 $\mib{r}_{i}$ indicates the position of the $i$-th impurity,
and $c_{\mib{r}\sigma}$ indicates the electron annihilation operator
at position $\mib{r}$.

We solve this system by the following procedure.
First, we apply the Hartree--Fock--Gor'kov approximation to $H_0$
by introducing the superconducting gap.
Then, we include the effect of impurity scattering on the superconducting
electrons in a framework of the perturbation theory concerning $U$,
called a self-consistent $T$-matrix approximation.
\cite{Schmitt,Hirschfeld1,Hirschfeld2,Hotta1,Hotta2,Balatsky}

\subsection{Hartree--Fock--Gor'kov approximation}

Let us first consider the mean-field approximation extended to the
superconducting state, i.e., the Hartree--Fock--Gor'kov approximation.
Then, we obtain
\begin{equation}
H^{\rm MF}_0 =\sum_{\mib{k}\sigma}\varepsilon_{\mib{k}}
c^{\dag}_{\mib{k}\sigma}c_{\mib{k}\sigma}
-\sum_{\mib{k}}(\Delta^*_{\mib{k}}
c_{\mib{-k}\downarrow}c_{\mib{k}\uparrow}
+\Delta_{\mib{k}}
c^{\dag}_{\mib{k}\uparrow}c^{\dag}_{\mib{-k}\downarrow}),
\end{equation}
where the gap function $\Delta_{\mib{k}}$ is defined by
\begin{equation}
\Delta_{\mib{k}}=-\sum_{\mib{k'}}V_{\mib{k},\mib{k'}}
\langle c_{\mib{-k'}\downarrow}c_{\mib{k'}\uparrow} \rangle.
\end{equation}
Here, $\langle \cdots \rangle$ denotes the operation to take
the thermal average using $H_0^{\rm MF}$.
Note that the Hartree--Fock terms are already included in
the electron energy $\varepsilon_{\mib{k}}$,
and the constant energy shift including the gap function is
ignored in $H_0^{\rm MF}$.

It is possible to solve $H_0^{\rm MF}$ by a couple of methods, but here,
we employ the Green's function technique,\cite{Allen}
not the Bogoliubov canonical transformation method.
Let us now define the Green's function $\mib{G}_0(\mib{k},i\omega_n)$ as
\begin{equation}
\mib{G}_0(\mib{k}, i \omega_n)=
-\int_0^{1/T} d\tau e^{i\tau \omega_n}
\langle \Psi_{\mib{k}} (\tau)  \Psi_{\mib{k}}^{\dag} (0) \rangle,
\end{equation}
where  the boldface character represents the matrix in Nambu space,
$T$ is temperature,
and $\omega_n$ denotes the fermion Matsubara frequency,
given by $\omega_n=\pi T (2n+1)$ with an integer $n$,
$\Psi_{\mib{k}} (\tau) =e^{\tau H_0} \Psi_{\mib{k}} e^{-\tau H_0}$.
We introduce the two-component field operator $\Psi_{\mib{k}}$ as
\begin{equation}
\Psi_{\mib{k}}=
\left(
\begin{array}{l}
c_{\mib{k}\uparrow} \\
c_{\mib{-k}\downarrow}^{\dag}
\end{array}
\right),~~
\Psi_{\mib{k}}^{\dag}=
(c_{\mib{k}\uparrow}^{\dag}, c_{\mib{-k}\downarrow}).
\end{equation}
After some algebraic calculations, we obtain $\mib{G}_0$ as
\begin{equation}
\mib{G}^{-1}_0(\mib{k}, i \omega_n)=i\omega_n \sigma_0-
\Delta_{\bf k}\sigma_1-\varepsilon_{\mib{k}}\sigma_3, 
\end{equation}
where $\sigma_0$ is a $2 \times 2$ unit matrix,
while $\sigma_1$ and $\sigma_3$ are respectively given by
\begin{equation}
\sigma_1 =
\left(
\begin{array}{cc}
0 & 1 \\
1 & 0
\end{array}
\right),~~
\sigma_3 =
\left(
\begin{array}{cc}
1 & 0 \\
0 & -1
\end{array}
\right).
\end{equation}

After performing the summation in terms of the Matsubara frequency,
we obtain the gap equation as
\begin{equation}
\Delta_{\mib{k}}=-\sum_{\mib{k'}}V_{\mib{k},\mib{k'}}
\frac{\Delta_{\mib{k'}}}{2E_{\mib{k'}}}
\tanh \frac{E_{\mib{k'}}}{2T}.
\end{equation}
To solve the gap equation, we assume the
separable-type attractive interaction given by
\begin{equation}
V_{\mib{k},\mib{k'}}\!=\!
\left\{
\begin{array}{ll}
\! -V \psi_{\mib{k}} \psi_{\mib{k'}} & \! |\varepsilon_{\mib{k}}|<\omega_{\rm c}
~{\rm and}~|\varepsilon_{\mib{k'}}|<\omega_{\rm c},\\
\! 0 & {\rm otherwise},
\end{array}
\right.
\end{equation}
where $V(>0)$ is the magnitude of the attraction
and the cut-off frequency $\omega_{\rm c}$ is
assumed to be much less than the Fermi energy.
Accordingly, the gap function is given by
\begin{equation}
\Delta_{\mib{k}}=\Delta \psi_{\mib{k}},
\end{equation}
where $\Delta$ denotes the magnitude of the gap and,
in general, it depends on temperature and
impurity concentration.

In the weak-coupling limit, we solve the gap equation to obtain
the superconducting transition temperature $T_{\rm c0}$
and the gap at absolute zero temperature, $\Delta_0(0)$,
for the case without impurities.
Then, we obtain the ratio as
\begin{equation}
\label{gapratio}
\frac{\Delta_0(0)}{T_{\rm c0}}=\pi e^{-\gamma-\eta},
\end{equation}
where $\gamma$ denotes the Euler's constant and $\eta$ is given by
\begin{equation}
\eta=\frac{\langle \psi^2_{\mib{k}} \log |\psi_{\mib{k}}| \rangle _{\rm FS}}
{\langle \psi^2_{\mib{k}} \rangle _{\rm FS}}.
\end{equation} 
Here ``FS'' indicates the abbreviation of Fermi surface and
$\langle \cdots \rangle_{\rm FS}$ denotes the operation to
take the average over the Fermi-surface curve.
For the anisotropic gap, $\eta$ becomes negative,
leading to the enhancement of the ratio.

\subsection{Self-consistent $T$-matrix approximation}

Next, we include the nonmagnetic impurity effect,
\cite{Schmitt,Hirschfeld1,Hirschfeld2,Hotta1,Hotta2,Balatsky}
which is considered through the self-energy $\mib{\Sigma}$.
The Green's function obeys the Dyson's equation
\begin{equation}
\mib{G}^{-1}(\mib{k},z)=\mib{G}_0^{-1}(\mib{k},z)
-\mib{\Sigma}(\mib{k},z),
\end{equation}
where we use $z$ for the frequency,
which is obtained by the analytic continuation of $z=i\omega_n$,
and the impurity self-energy $\mib{\Sigma}$
should be expressed as
\begin{equation}
\mib{\Sigma}=\Sigma_0\sigma_0+\Sigma_1\sigma_1+\Sigma_3\sigma_3.
\end{equation}
Then, we obtain 
\begin{equation}
\mib{G}^{-1}(\mib{k},{\tilde z})={\tilde z}\sigma_0-
{\tilde \Delta}_{\bf k}\sigma_1-{\tilde \varepsilon}_{\bf k}\sigma_3,
\end{equation}
where 
\begin{equation}
\label{sol0}
{\tilde z}=z-\Sigma_0,~~
{\tilde \Delta}_{\bf k}=\Delta_{\bf k}+\Sigma_1,~~
{\tilde \varepsilon}_{\bf k}=\varepsilon_{\bf k}+\Sigma_3.
\end{equation}
Hereafter, we simply ignore $\Sigma_3$, since this term is not essentially
important in the following discussion.
In fact, the correction to one-electron energy is considered to be
included at the level of Hartree--Fock approximation.

Let us evaluate $\mib{\Sigma}$ in a single-site approximation given by
\begin{equation}
\mib{\Sigma}({\tilde z})=n_{\rm imp} \mib{T}({\tilde z}),
\end{equation}
where $n_{\rm imp}$ indicates the impurity concentration
and $\mib{T}$ denotes the $T$-matrix for impurity scattering.
The $T$-matrix obeys the equation given by
\begin{equation}
\mib{T}({\tilde z})=\mib{U}+\mib{ U}\sum_{\mib{k}} \mib{G}(\mib{k},{\tilde z})
\mib{T}({\tilde z}),
\end{equation}
where $\mib{U}=U\sigma_3$ .
In the unitarity limit, i.e., $UN_0(0) \gg 1$, where $N_0(0)$ denotes
the DOS at the Fermi level in the normal state,
we obtain
\begin{equation}
\label{sol1}
\Sigma_0=-\alpha \frac{g_0}{g_0^2-g_1^2}, ~~
\Sigma_1=\alpha \frac{g_1}{g_0^2-g_1^2},
\end{equation}
where $\alpha$ is a pair-breaking parameter given by
\begin{equation}
\label{sol2}
\alpha=\frac{n_{\rm imp}}{\pi N_0(0)}.
\end{equation}
The averaged Green's functions $g_0$ and $g_1$ are respectively given by
\begin{equation}
\label{sol3}
g_0=- \left\langle \frac{\tilde z}{
\sqrt {{\tilde \Delta}_{\bf k}^2-{\tilde z}^2}} \right\rangle _{\! \! \! {\rm FS}}, ~~
g_1=- \left\langle \frac{{\tilde \Delta}_{\bf k}}
{\sqrt {{\tilde \Delta}_{\bf k}^2-{\tilde z}^2}} \right\rangle _{\! \! \! {\rm FS}}.
\end{equation}
The normalized DOS $N(z)$ in the superconducting state
affected by impurity scattering is given by
\begin{equation}
N(z)=-{\rm Im} g_0.
\end{equation}
Note that $N(z)$ is normalized by $N_0(0)$ and $N(\infty)=1$.

In this paper, we evaluate the nuclear magnetic relaxation rate $T_1^{-1}$
in the superconducting state,
since it is sensitive to the change in DOS in the low-energy region.
When we define $R=(T_1T)^{-1}$, we discuss the temperature dependence
of the ratio $R_{\rm s}(T)/R_{\rm n}$, where the subscripts ``s'' and ``n''
indicate the superconducting and normal states, respectively.
By following the Bardeen--Cooper--Schrieffer theory,\cite{BCS}
we obtain $R_{\rm s}(T)/R_{\rm n}$ as 
\begin{equation}
\label{t1t}
\frac{R_{\rm s}(T)}{R_{\rm n}}=
2\int_0^{\infty}dz \left(-\frac{\partial f}{\partial z}\right)
\left[N^2(z) + M^2(z) \right], 
\end{equation}
where $f=1/(e^{z/T}+1)$ and $M(z)$ is given by
\begin{equation}
M(z)=-{\rm Im} g_1.
\end{equation}
We are also interested in the temperature dependence of
superfluid density, $\rho_{\rm s}(T)$,
which is evaluated from results of the penetration depth experiment.
It is given by
\begin{equation}
\label{rhos}
\frac{\rho_{\rm s}(T)}{\rho_{\rm n}}=
1-2\int_0^{\infty}dz \left(-\frac{\partial f}{\partial z}\right) N(z),
\end{equation}
where $\rho_{\rm n}$ indicates the electron density in the normal state.

Here, we comment on the assumption
when we will perform numerically the above integrals.
It is assumed that the ratio of $T_{\rm c}$ to the gap at $T=0$
is not affected by impurity scattering.
Namely, we use the relation
$\Delta(0)/T_{\rm c}=\Delta_0(0)/T_{\rm c0}$
and the ratio is given by Eq.~(\ref{gapratio}).

%%%%%%%%%%%%%%%%%%%%%%% fig. 1 %%%%%%%%%%%%%%%%%%%%%%%%%%%
\begin{figure}[t]
\centering
\includegraphics[width=8.0truecm]{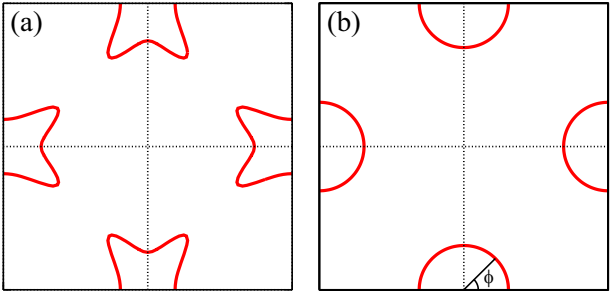}
\caption{(Color online) (a) Fermi-surface curves for LaO$_{1-x}$F$_x$BiS$_2$
with $x=0.3$ obtained on the basis of the tight-binding model.
(b) Model Fermi-surface curves used in this paper.
The electron density is set as the value corresponding to $x=0.3$.
}
\end{figure}
%%%%%%%%%%%%%%%%%%%%%%%%%%%%%%%%%%%%%%%%%%%%%%%%%%%%%

\subsection{Fermi-surface structure and gap functions}

The canonical model for BiS$_2$-based layered superconductors
was proposed by Usui and coworkers,\cite{Usui,Suzuki},
just after the discovery of BiS$_2$-based layered superconductors.
The minimal model to describe the electronic structure of
the BiS$_2$ layer is the two-band Hamiltonian composed of
Bi 6$p_x$ and 6$p_y$ orbitals on the two-dimensional square lattice.
In Fig.~1(a), we show the Fermi-surface curves for $x=0.3$
of the minimal model.
We point out a characteristic issue that
pocketlike disconnected Fermi-surface curves appear around
$\mib{k}=(\pm \pi,0)$ and $(0, \pm \pi)$.
Note that for $x<0.52$, we find the Fermi-surface curve originating
from the lower-energy band.\cite{Usui,Suzuki,Aga}
Thus, for $x=0.3$, we consider effectively only the band forming
the Fermi-surface curves for the appearance of superconductivity.

In order to concentrate on the impurity effect in this case,
we simplify the Fermi-surface structure by maintaining its topology.
As shown in Fig.~1(b), we consider four semicircle Fermi-surface curves
at around $\mib{k}=(\pm \pi, 0)$ and $(0, \pm \pi)$,
in order to reproduce the pocketlike disconnected
Fermi-surface structure of the effective model for BiS$_2$-based layered materials.
For the operation to take the average over the Fermi-surface curve,
it is useful to define the angle $\phi$ to specify the position on the Fermi-surface
curve, as shown in Fig.~1(b).

In this paper, we consider both extended $s$- and $d$-wave gap functions.
Note that we do not consider the $s$-wave gap here, since
it is well known that nonmagnetic impurities do not affect it.
When we consider the extended $s$-wave gap
$\Delta_{\mib{k}} \propto \cos k_x + \cos k_y$,
the nodal lines cross the Fermi-surface curves, as shown in Fig.~2(a).
When we use the angle $\phi$ defined in Fig.~1(b),
the gap is well approximated by $\Delta(\phi)=\Delta \cos (2\phi)$
on the Fermi-surface curves.
This is essentially the same function as the $d$-wave gap
on the large Fermi-surface curve with the center at the $\Gamma$ point
for high-$T_{\rm c}$ cuprates.
Note, however, that the extended $s$-wave gap is allowed
to have a constant component.
Thus, in general, the extended $s$-wave gap is written as
\begin{equation}
\label{esgap}
\Delta(\phi)=\Delta [p + \cos (2\phi)],
\end{equation}
where $\Delta$ in this case is the gap of the anisotropic part and
$p$ denotes the ratio of isotropic to anisotropic gaps.
In the present Fermi-surface structure, the node positions move
from $\phi=\pi/4$ and $3\pi/4$ for $p=0$
to $\phi=\pi/2$ for $p=1$.
For $p>1$, the nodes do not appear on the Fermi-surface curves.
The impurity effect should be different depending on the gap ratio $p$.
This point will be discussed in detail later.

%%%%%%%%%%%%%%%%%%%%%%% Fig. 2 %%%%%%%%%%%%%%%%%%%%%%%%%%%
\begin{figure}[t]
\centering
\includegraphics[width=8.0truecm]{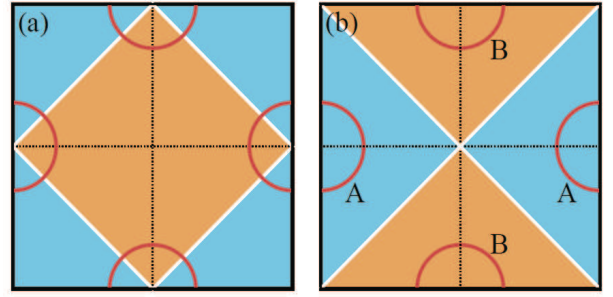}
\caption{(Color online)
(a) Nodal extended $s$-wave gap.
The blue and orange regions denote $\Delta_{\mib{k}}<0$
and $\Delta_{\mib{k}}>0$, respectively.
The white lines between these regions denote the gap nodes,
which cross the Fermi-surface curves.
(b) Nodeless $d$-wave gap.
The Fermi-surface curves in the blue and orange regions are
denoted by A and B, respectively.
}
\end{figure}
%%%%%%%%%%%%%%%%%%%%%%%%%%%%%%%%%%%%%%%%%%%%%%%%%%%%%

Next, we consider the $d$-wave gap.
In sharp contrast to the case with a large Fermi-surface curve
with the center at the $\Gamma$ point,
the present Fermi-surface curves do $not$ cross the lines of
$k_x=\pm k_y$.
Thus, when we assume the $d$-wave gap,
$\Delta_{\mib{k}} \propto \cos k_x - \cos k_y$,
the nodes do not appear on the Fermi-surface curves,
as shown in Fig.~2(b).
In this sense, it can be called the nodeless $d$-wave gap.
Note, however, that the gap has the same sign on the Fermi-surface curve,
while the sign change occurs between the gap functions on the different pocketlike
Fermi-surface curves.
Thus, when we consider the gap on the pocketlike Fermi-surface curve,
it looks like a simple $s$-wave gap at first glance,
but the average of the gap over the whole Fermi-surface curves becomes zero.
This fact has a remarkable impact on the impurity effect on
nodeless $d$-wave superconductors.
In the following calculations, we consider the nodeless $d$-wave gap as
\begin{equation}
\label{dgap}
\Delta_{\mib{k}}=
\left\{
\begin{array}{ll}
-\Delta & {\rm on~the~FS~A}, \\
+\Delta & {\rm on~the~FS~B},
\end{array}
\right.
\end{equation}
where $\Delta$ is the magnitude of the gap,
and the Fermi-surface curves A and B are defined in Fig.~2(b).
Note here that for simplicity, we ignore the $\mib{k}$ dependence
of the gap on the Fermi-surface curves,
but it is easy to check the validity of this approximation.

\subsection{Reduction in $T_{\rm c}$}

Before showing our calculation results
concerning the impurity effect on physical quantities,
let us briefly discuss the reduction in the transition
temperature $T_{\rm c}$ due to nonmagnetic impurity scattering.

First, we consider the extended $s$-wave gap.
Except for the case of $p=0$, the well-known Abrikosov$-$Gor'kov formula
for the $T_{\rm c}$ reduction \cite{AG} cannot be simply used,
since there exists a constant component $p$.
As pointed out by Tsuneto,\cite{Tsuneto}
it is necessary to consider the change in the pairing interaction
due to the nonmagnetic impurity scattering.
After some algebraic calculations, we obtain the generalized formula
\begin{equation}
\log \left( \frac{T_{\rm c0}}{T_{\rm c}} \right)
= \beta \left[
\psi\left( \frac{1}{2}+\frac{\alpha}{2\pi T_{\rm c}} \right)
-\psi\left( \frac{1}{2} \right)
\right],
\end{equation}
where $T_{\rm c0}$ indicates the superconducting transition temperature
without nonmagnetic impurity,
$\psi$ is the di-gamma function, and $\beta$ is given by
\begin{equation}
\beta=1-\frac{\langle \Delta_{\mib{k}} \rangle^2_{\rm FS}}
{\langle \Delta_{\mib{k}}^2 \rangle_{\rm FS}}.
\end{equation}
In the case of the simple $s$-wave gap, $\Delta_{\mib{k}}$ is constant
and thus, $\beta=0$, indicating that $T_{\rm c}$ is not affected at all
by the nonmagnetic impurity.
On the other hand, for the $d$-wave gap,
since $\langle \Delta_{\mib{k}} \rangle_{\rm FS}=0$,
we obtain $\beta=1$, leading to the well-known Abrikosov--Gor'kov formula.

For the present extended $s$-wave gap Eq.~(\ref{esgap}),
we can take the average over the Fermi-surface curves.
Then, we obtain $\beta$ as
\begin{equation}
\beta=\frac{1}{2p^2+1}.
\end{equation}
In Fig.~3(a), we show $T_{\rm c}/T_{\rm c0}$
as functions of $\zeta$ at various values of $p$,
where $\zeta$ is defined as
\begin{equation}
\zeta=\frac{\alpha}{2\pi T_{\rm c0}}.
\end{equation}
Here, we briefly explain the choice of $p$.
From the form of the gap $p+\cos(2\phi)$,
we immediately recognize that the node of the gap disappears at $p=1$.
We also note that the pure anisotropic case, $p=0$, is exceptional.
Namely, it is necessary to consider four regions 
as $p=0$, $0<p<1$, $p=1$, and $p>1$.
Thus, we show the results for four values of $p$ as $p=0$, $0.5$, $1$, and $1.5$.

In the curve for $p=0$, i.e., $\beta=1$, $T_{\rm c}$ becomes zero
at $\zeta=\zeta_{\rm c}$,
where $\zeta_{\rm c}$ denotes the critical value of $\zeta$,
at which $T_{\rm c}$ becomes zero.
By using the asymptotic form of the di-gamma function $\psi$, given by
\begin{equation}
\psi\left( \frac{1}{2}+x \right)-\psi\left( \frac{1}{2} \right) \approx \log (4e^{\gamma} x),
\end{equation}
for $x \gg 1$, we obtain $\zeta_{\rm c}=e^{-\gamma}/4=0.14$.

When the isotropic gap exists for $p>0$,
the $T_{\rm c}$ reduction is gradual for large $p$ values
and there are no critical $\zeta$ value.
The anisotropic part of the gap is washed out by
the non-magnetic impurity scattering,
indicating that the isotropic gap remains. 
This point has already been emphasized by Tsuneto.\cite{Tsuneto}

%%%%%%%%%%%%%%%%%%%%%%% Fig. 3 %%%%%%%%%%%%%%%%%%%%%%%%%%%
\begin{figure}[t]
\centering
\includegraphics[width=8.0truecm]{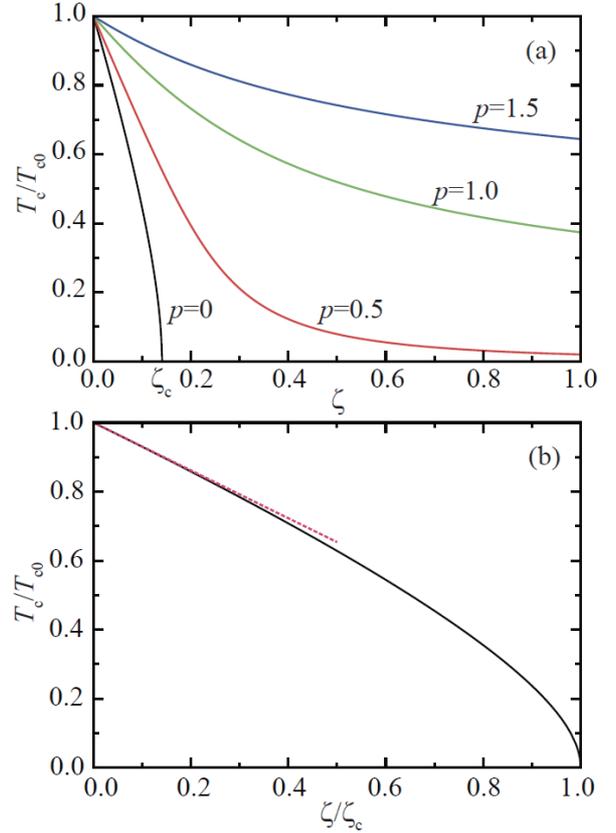}
\caption{(Color online) 
(a)  $T_{\rm c}/T_{\rm c0}$ vs $\zeta$ for the extended $s$-wave gap with
$p=0$, $0.5$, $1.0$, and $1.5$.
Note that in this case, except for the case of $p=0$, there exists no critical value of $\zeta$.
(b) $T_{\rm c}/T_{\rm c0}$ as a function of $\zeta/\zeta_{\rm c}$
for nodeless $d$-wave gap.
Note that the red broken line denotes Eq.~(\ref{tcreducapp}).
}
\end{figure}
%%%%%%%%%%%%%%%%%%%%%%%%%%%%%%%%%%%%%%%%%%%%%%%%%%%%%

Next, we consider the case of the nodeless $d$-wave gap.
In a previous study on the nonmagnetic impurity
effect in nodal $d$-wave superconductors,\cite{Hotta1}
the reduction in $T_{\rm c}$ is given by the
Abrikosov--Gor'kov formula,
\begin{equation}
\log \left( \frac{T_{\rm c0}}{T_{\rm c}} \right)
= \psi\left( \frac{1}{2}+\frac{\alpha}{2\pi T_{\rm c}} \right)
-\psi\left( \frac{1}{2} \right).
\end{equation}
Also, in the nodeless $d$-wave case,
this Abrikosov-Gor'kov formula is available,
since the average of the gap over the Fermi-surface curve vanishes.

In Fig.~3(b), we show the curve for $T_{\rm c}/T_{\rm c0}$
as a function of $\zeta/\zeta_{\rm c}$.
This is essentially the same as the curve for $p=0$ in Fig.~3(a).
We consider the approximate expression for $\zeta/\zeta_{\rm c} \ll 1$.
With the use of the expansion formula of the di-gamma function for $x \ll 1$,
given by
\begin{equation}
\psi\left( \frac{1}{2}+x \right)-\psi\left( \frac{1}{2} \right) \approx \frac{\pi^2}{2}x,
\end{equation}
$T_{\rm c}/T_{\rm c0}$ is well approximated as
\begin{equation}
\label{tcreducapp}
\frac{T_{\rm c}}{T_{\rm c0}}=1-\frac{\pi^2}{8e^{\gamma}}\frac{\zeta}{\zeta_{\rm c}},
\end{equation}
in the case of $T_{\rm c0}-T_{\rm c} \ll T_{\rm c0}$.
This is plotted by the red line, which well agrees with the Abrikosoz-Gor'kov formula
in the region of $\zeta/\zeta_{\rm c} \ll 1$.

\section{Calculation Results}

\subsection{Extended $s$-wave gap}

\subsubsection{Self-consistent equations}

Let us briefly explain the equations for the extended $s$-wave gap.
For $\zeta=0$, $N(\omega)$ is evaluated as
\begin{equation}
N(\omega) = {\rm Im} \left\langle \frac{\omega}{
\sqrt{[p+\cos (2\phi)]^2-\omega^2}} \right\rangle _{\! \! \! {\rm FS}}.
\end{equation}
For the case with nonmagnetic impurities,
we solve the self-consistent equations Eqs.~(\ref{sol0}), (\ref{sol1}), (\ref{sol2}),
and (\ref{sol3}) for the extended $s$-wave gap Eq.~(\ref{esgap}).
Here, we note that $g_1$ does not vanish in general.
Note also that the effect of the $g_1$ term has a significant contribution
to the isotropic part, while the anisotropic part is not affected at all
by the nonmagnetic impurity scattering.
To obtain $g_0$ and $g_1$ with nonmagnetic impurities,
we rewrite the self-consistent equations as
\begin{equation}
{\tilde \omega} =\omega+\xi \frac{g_0}{g_0^2-g_1^2},
{\tilde p} =p+\xi \frac{g_1}{g_0^2-g_1^2},
\end{equation}
where $\omega=z/\Delta$, ${\tilde \omega}={\tilde  z}/\Delta$,
and $\xi=\alpha/\Delta$.

Green's functions $g_0$ and $g_1$ are given by
\begin{equation}
\label{g0g1}
\begin{split}
g_0 &=-\frac{1}{\pi} \int_0^{\pi} d\phi \frac{{\tilde \omega}}
{\sqrt{[{\tilde p}+\cos(2\phi)]^2-{\tilde \omega}^2}},\\
g_1 &=-\frac{1}{\pi} \int_0^{\pi} d\phi  \frac{{\tilde p}+\cos(2\phi)}
{\sqrt{[{\tilde p}+\cos(2\phi)]^2-{\tilde \omega}^2}},
\end{split}
\end{equation}
respectively.

We solve the above equations self-consistently
concerning ${\tilde \omega}$ and ${\tilde p}$.
Note that for the anisotropic gap with $p=0$,
we easily find the solution of ${\tilde p}=g_1=0$.
In this case, it is sufficient to solve the self-consistent equation
concerning ${\tilde \omega}$,
as in the case of the nodeless $d$-wave gap.

Throughout the calculation of the DOS, since the energy unit is set as $\Delta$,
we define $\xi$ as $\xi=\alpha/\Delta$, but it is different from $\zeta$.
The energy unit $\Delta$ is the solution of the gap equation and
it depends on the temperature and the impurity concentration,
as mentioned above.
In the calculation of $R_{\rm s}/R_{\rm n}$ and $\rho_{\rm s}/\rho_{\rm n}$,
we will consider explicitly the temperature dependence of $\Delta$
for a certain value of $\zeta$.

%%%%%%%%%%%%%%%%%%%%%% Fig. 4 %%%%%%%%%%%%%%%%%%%%%%%%%%%
\begin{figure}[t]
\centering
\includegraphics[width=8.0truecm]{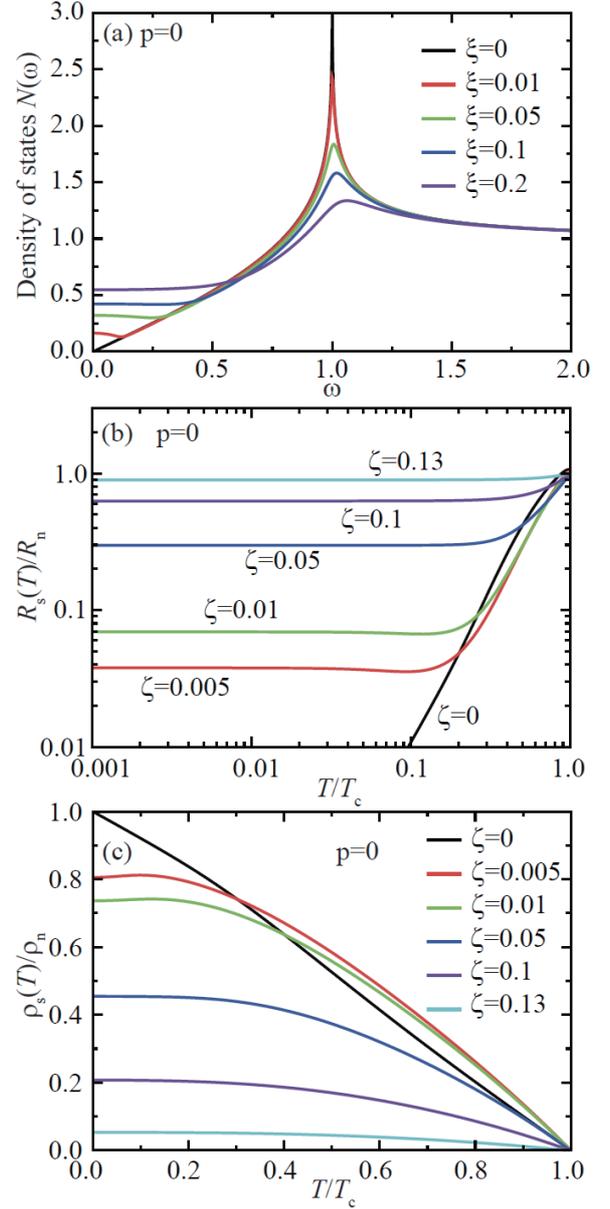}
\caption{(Color online) Results for extended $s$-wave gap with $p=0$.
(a) $N(\omega)$ vs $\omega$.
(b) $R_{\rm s}(T)/R_{\rm n}$ vs $T/T_{\rm c}$ in a logarithmic scale.
(c) Superfluid density $\rho_{\rm s}/\rho_{\rm n}$ vs $T/T_{\rm c}$.
}
\end{figure}
%%%%%%%%%%%%%%%%%%%%%%%%%%%%%%%%%%%%%%%%%%%%%%%%%%%%%

\subsubsection{Results for $p=0$}

In Fig.~4(a), we show the results of the DOS for $p=0$.
First, we consider the case of $\xi=0$.
After some algebraic calculations, we obtain the DOS as
\begin{equation}
N(\omega)=\frac{2}{\pi}\omega K(\omega),
\end{equation}
for $0 \le \omega <1$, while for $ \omega>1$, we obtain
\begin{equation}
N(\omega)=\frac{2}{\pi}K\left( \frac{1}{\omega} \right),
\end{equation}
where $K(k)$ is the complete elliptic integral of the first kind
defined as
\begin{equation}
K(k)=\int_0^{\pi/2} d\theta \frac{1}{\sqrt{1-k^2 \sin ^2 \theta}}.
\end{equation}
We remark that $N(\omega)=\omega$ near the Fermi level,
which is essentially the same as the $d$-wave gap
on the Fermi-surface curves with the center at the $\Gamma$ point.
We find the logarithmic divergence at $\omega=1$,
which is characteristic of the two-dimensional case.\cite{Hotta3}
For $\xi>0$, we find the almost constant DOS near the Fermi
surface and the logarithmic divergence is smeared out.
The behavior due to the resonant impurity scattering is also
the same as that for the nodal $d$-wave gap.\cite{Hotta1,Hotta2}

In Figs.~4(b) and 4(c), we show the results of the nuclear magnetic relaxation rate
and superfluid density for the extended $s$-wave gap with $p=0$.
For $\zeta=0$, owing to the effect of the nodes on the Fermi-surface curves,
we observe the power-law behavior of $T_1^{-1} \propto T^3$
and $1-\rho_s/\rho_n\propto T$,
characteristic of the $d$-wave superconductors.
When we dope nonmagnetic impurities,
a finite DOS appears at the low-energy region,
leading to the revival of the Korringa law,
i.e.,  $T_1T$=constant,
and the $s$-wave-like constant behavior of $\rho_s/\rho_n$
at low temperatures.
Note that $\rho_s/\rho_n$ is reduced from unity
at $T=0$, since part of $\rho_s$ is changed to $\rho_n$
owing to the pair-breaking effect.

%%%%%%%%%%%%%%%%%%%%%% Fig. 5 %%%%%%%%%%%%%%%%%%%%%%%%%%%
\begin{figure}[t]
\centering
\includegraphics[width=8.0truecm]{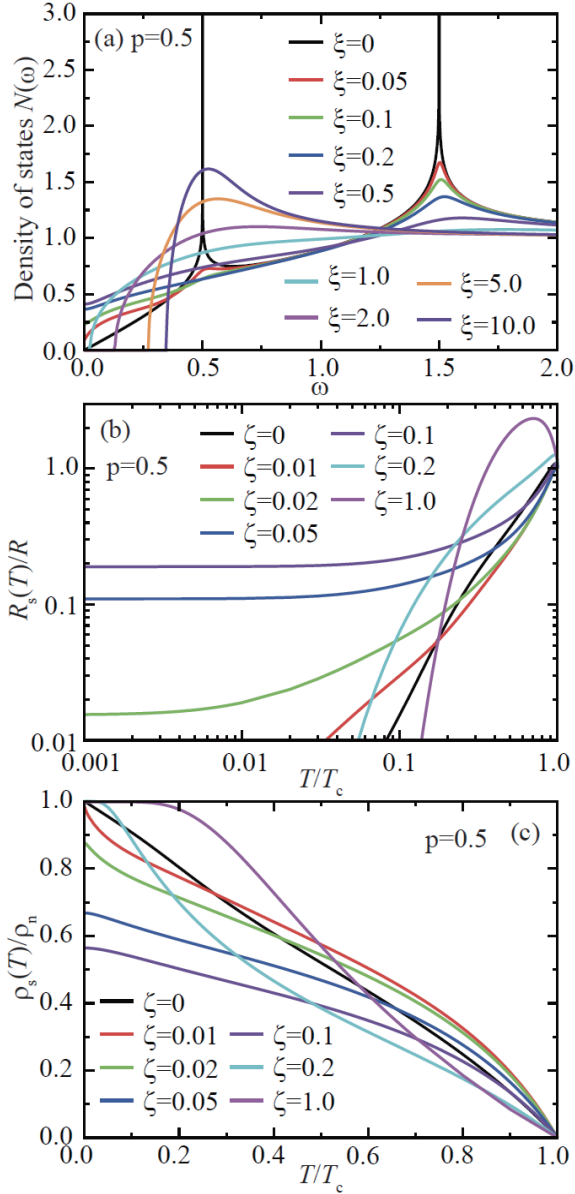}
\caption{(Color online) Results for extended $s$-wave gap with $p=0.5$.
(a) $N(\omega)$ vs $\omega$.
(b) $R_{\rm s}(T)/R_{\rm n}$ vs $T/T_{\rm c}$ in a logarithmic scale.
(c) Superfluid density $\rho_{\rm s}/\rho_{\rm n}$ vs $T/T_{\rm c}$.
}
\end{figure}
%%%%%%%%%%%%%%%%%%%%%%%%%%%%%%%%%%%%%%%%%%%%%%%%%%%%%

\subsubsection{Results for $p=0.5$}

In Fig.~5(a), we show the results of the DOS for $p=0.5$.
For both $0 \le \omega <1-p$ and $\omega > 1+p$,
we obtain the DOS as
\begin{equation}
\label{dos1}
N(\omega)=\frac{2}{\pi}\frac{\omega}{\sqrt{(1+\omega)^2-p^2}}
K(A_{\omega}),
\end{equation} 
where $A_{\omega}$ is given by
\begin{equation}
\label{aomega}
A_{\omega}= \sqrt{\frac{4\omega}{(1+\omega)^2-p^2}}.
\end{equation}
For $1-p < \omega < 1+p$, we obtain
\begin{equation}
\label{dos2}
N(\omega)=\frac{\sqrt{\omega}}{\pi}
K\left( \frac{1}{A_{\omega}} \right).
\end{equation}
For $0<p<1$, since the nodes still exist on the Fermi-surface curves,
the DOS is in proportion to $\omega$ near the Fermi level.
As denoted by $N(\omega)=\omega/\sqrt{1-p^2}$ near $\omega=0$,
the slope becomes steep in comparison with that of $p=0$.
We observe two anomalies at $\omega=1-p$ and $1+p$
in the DOS.

When we dope nonmagnetic impurities, both anomalies are washed out
and the finite DOS appears at $\omega=0$.
However, in sharp contrast to the case of $p=0$,
we do not observe the almost constant behavior near the Fermi level.
For $\xi>0.5$, the DOS at $\omega=0$ begins to decrease and eventually,
it becomes zero at $\xi \approx 1$.
Then, the gap opens at the low-energy region, since the anisotropic part
is washed out by non-magnetic impurity scattering.
If we increase $\xi$ further up to an unrealistically large value,
the DOS approaches the form of $\omega/\sqrt{\omega^2-p^2}$,
where the size of the effective gap is $p \Delta$.
Note again that such a situation is realized only mathematically,
since it is necessary to dope huge amounts of impurities.

Next, we show the results of nuclear magnetic relaxation rate
and superfluid density for the extended $s$-wave gap with $p=0.5$
in Figs.~5(b) and 5(c).
Since the nodes of the gap still exist on the Fermi-surface curves,
we observe the power-law behavior of $T_1^{-1} \propto T^3$
and $1-\rho_s/\rho_n \propto T$ for $\zeta=0$,
although the slope is different from that in the case of $p=0$.
For a small $\zeta$, owing to the finite DOS at the Fermi level,
we observe the almost constant value of $R_{\rm s}/R_{\rm n}$
at low temperatures, as shown in Fig.~5(b).
Note, however, that the flat region is apparently narrow in comparison
with the case of $p=0$, since the DOS is not constant
at low-energy regions for $p=0.5$.
As shown in Fig.~5(c),
this effect can be clearly found in $\rho_s/\rho_n$ for $\zeta < 0.2$,
which is not considered to be constant at low temperatures,
although $\rho_s/\rho_n$ at $T=0$ is reduced from unity.
When we further increase the value of $\zeta$, as mentioned in
the discussion on the DOS,
a finite gap begins to open near the Fermi level.
This effect appears in the $s$-wave-like behavior of
$R_{\rm s}/R_{\rm n}$ and $\rho_{\rm s}/\rho_{\rm n}$ for large $\zeta$.
In particular, for $\zeta=1$, we observe a large coherence peak
just below $T_{\rm c}$.

%%%%%%%%%%%%%%%%%%%%%% Fig. 6 %%%%%%%%%%%%%%%%%%%%%%%%%%%
\begin{figure}[t]
\centering
\includegraphics[width=8.0truecm]{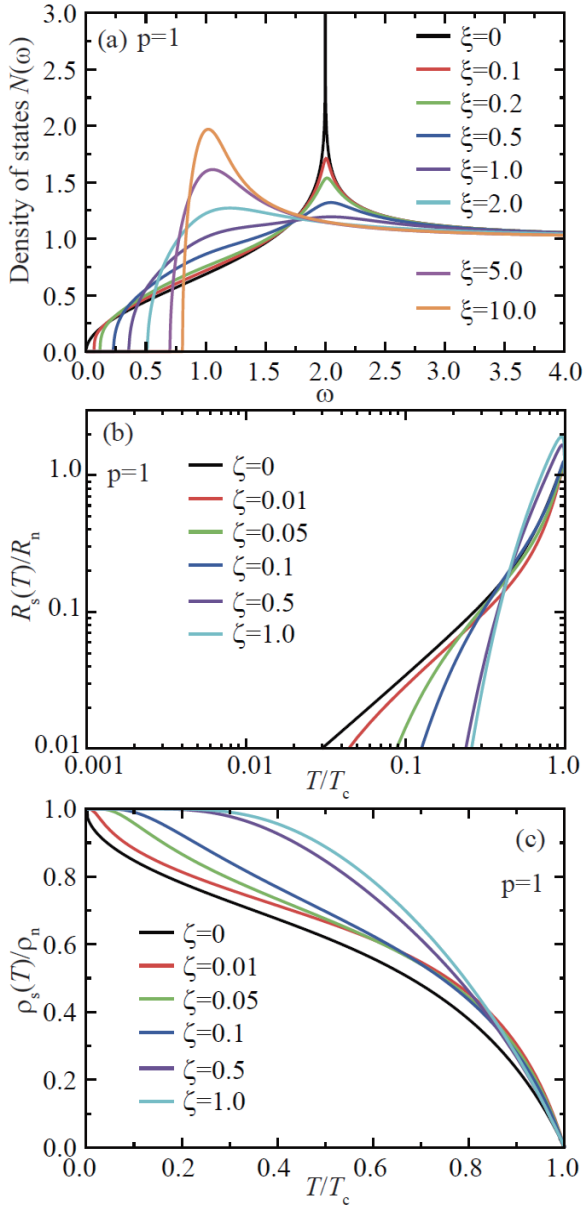}
\caption{(Color online) Results for extended $s$-wave gap with $p=1$.
(a) $N(\omega)$ vs $\omega$.
(b) $R_{\rm s}(T)/R_{\rm n}$ vs $T/T_{\rm c}$ in a logarithmic scale.
(c) Superfluid density $\rho_{\rm s}/\rho_{\rm n}$ vs $T/T_{\rm c}$.
}
\end{figure}
%%%%%%%%%%%%%%%%%%%%%%%%%%%%%%%%%%%%%%%%%%%%%%%%%%%%%

\subsubsection{Results for $p=1$}

In Fig.~6(a),  we show the results of the DOS for $p=1$.
For $\xi=0$, we obtain the analytic form of the DOS.
In the region of $0 \le \omega <2$, we obtain
\begin{equation}
N(\omega)=\frac{\sqrt{\omega}}{\pi}
K\left( \frac{\sqrt{\omega+2}}{2} \right),
\end{equation}
while for $\omega>2$, the DOS is given by
\begin{equation}
N(\omega)=\frac{2}{\pi} \sqrt{\frac{\omega}{\omega+2}}
K\left (\frac{2}{\sqrt{\omega+2}} \right).
\end{equation}
Note that near the Fermi level, $N(\omega)$ is proportional to
$\sqrt{\omega}$ and the logarithmic anomaly appears at $\omega=2$.

When we increase the value of $\xi$,
the logarithmic anomaly at $\omega=2$ is smeared
by the nonmagnetic impurity scattering.
Note that for $p=1$, the finite DOS does $not$ appear
at the Fermi level owing to the impurity scattering,
in sharp contrast to the case of $p=0$.
Rather, the finite gap begins to open near the Fermi level.
The size of the gap monotonically increases with the increase in $\xi$.
As expected from the curve for $\xi=10$ in Fig.~6(a),
the DOS asymptotically approaches
$\omega/\sqrt{\omega^2-1}$
with the effective gap of $\Delta$
for $\xi \rightarrow \infty$

In Figs.~6(b) and 6(c), we show the results of  $R_{\rm s}(T)/R_{\rm n}$
and $\rho_{\rm s}(T)/\rho_{\rm n}$ for $p=1$.
Note that the DOS near the Fermi level
behaves as $N(\omega) \propto \sqrt{\omega}$ for $p=1$,
since the node exists only at $\phi=\pi/2$ on the Fermi-surface curve.
For $\xi=0$, owing to this low-energy behavior of the DOS,
we find $(T_1T)^{-1} \propto T$ and $1-\rho_s/\rho_n \propto \sqrt{T}$.
When $\xi$ increases with impurity doping,
as mentioned in the discussion on the DOS,
a finite gap immediately opens at the Fermi level.
Thus, even for a small $\xi$, we observe the reappearance of
the $s$-wave-like behavior for both $(T_1T)^{-1}$ and $\rho_s/\rho_n$.

%%%%%%%%%%%%%%%%%%%%%% Fig. 7 %%%%%%%%%%%%%%%%%%%%%%%%%%%
\begin{figure}[t]
\centering
\includegraphics[width=8.0truecm]{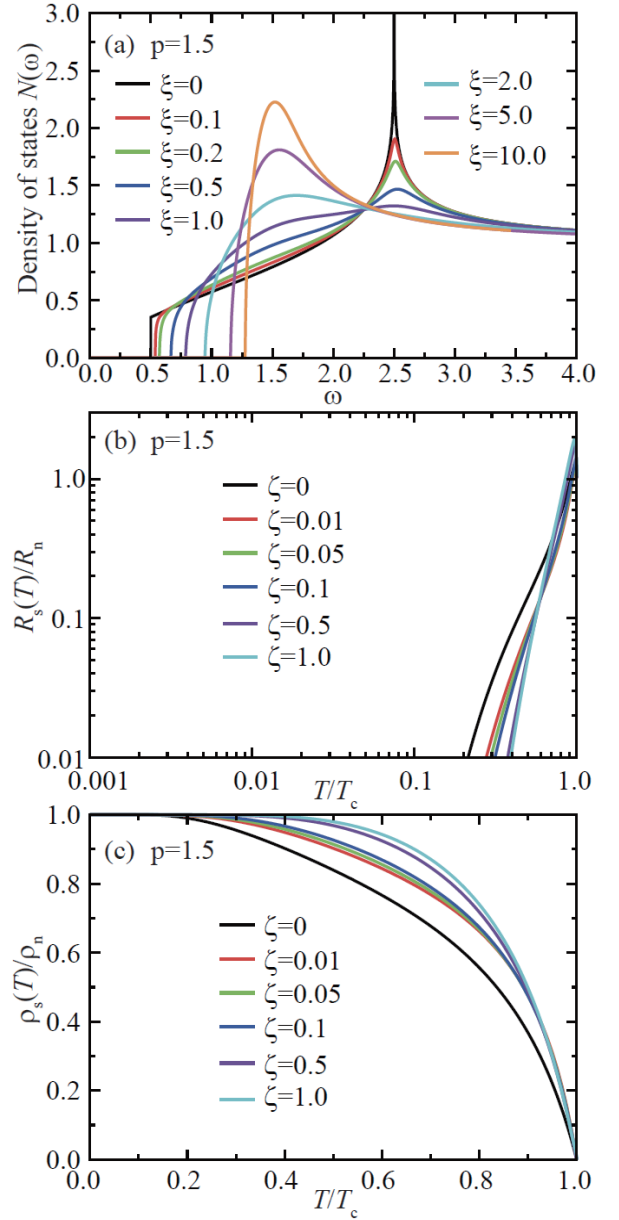}
\caption{(Color online) Results for extended $s$-wave gap with $p=1.5$.
(a) $N(\omega)$ vs $\omega$.
(b) $R_{\rm s}(T)/R_{\rm n}$ vs $T/T_{\rm c}$ in a logarithmic scale.
(c) Superfluid density $\rho_{\rm s}/\rho_{\rm n}$ vs $T/T_{\rm c}$.
}
\end{figure}
%%%%%%%%%%%%%%%%%%%%%%%%%%%%%%%%%%%%%%%%%%%%%%%%%%%%%

\subsubsection{Results for $p=1.5$}

In Fig.~7(a), the results of the DOS for $p=1.5$ are shown.
For $\xi=0$, in the region of $0<\omega<p-1$, we easily obtain
$N(\omega)=0$ owing to the gap with the magnitude of $p-1$.
For $p-1< \omega < p+1$, the DOS is given by Eq.~(\ref{dos2}),
while for $\omega>1+p$, the DOS is given by Eq.~(\ref{dos1}).
The gap with the size of $(p-1)\Delta$ opens near the Fermi level,
since the anisotropic part is relatively smaller than the isotropic gap.
The anomaly at $\omega=p-1$ appears as the gap edge, while
the logarithmic divergence is found at $\omega=p+1$.
For $\xi>0$, the anomalies are smeared out and the gap edge is
found to be shifted toward the isotropic gap edge $\omega=p=1.5$.
Since the low-energy part of the DOS is not so
affected by the impurity scattering, the impurity
effect is expected to be less sensitive for $p>1$.

In Figs.~7(b) and 7(c), we depict  $R_s/R_n$ and $\rho_s/\rho_n$,
respectively, for $p=1.5$.
As already emphasized in the discussion on the DOS for $p>1$,
the gap with the magnitude of $(p-1)\Delta$ exists even at $\zeta=0$.
With impurity doping, the anisotropic part of the gap is gradually washed out 
and then the gap size is changed to $p\Delta$ for a large $\zeta$.
Namely, as long as we concentrate on the low-energy region,
we always expect the simple $s$-wave-like behavior
in physical quantities.
In fact, for both $R_s/R_n$ and $\rho_s/\rho_n$, the changes in the temperature
dependence are not significant when we increase $\zeta$.
In this sense, the temperature dependence is insensitive to
the nonmagnetic impurity for $p>1$.

%%%%%%%%%%%%%%%%%%%%%% Fig. 8 %%%%%%%%%%%%%%%%%%%%%%%%%%%
\begin{figure}[t]
\centering
\includegraphics[width=8.0truecm]{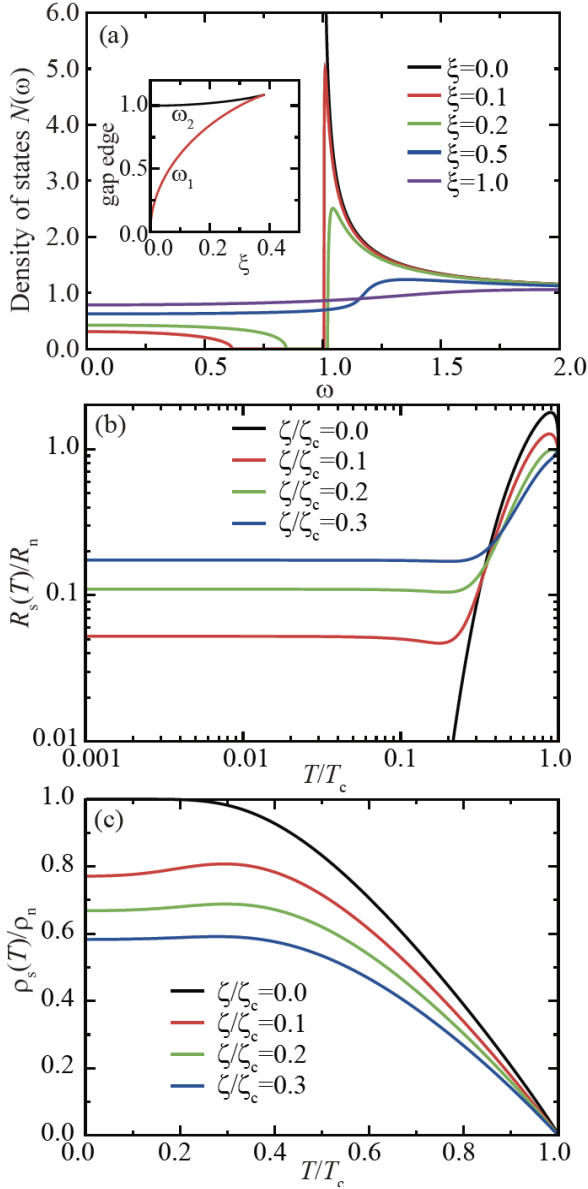}
\caption{(Color online) Results for nodeless $d$-wave gap.
(a) $N(\omega)$ vs $\omega$.
In the inset, we show gap edges vs $\xi$.
The upper and lower edges are defined as $\omega_2$ and $\omega_1$, respectively.
Note that the gap becomes zero at $\xi=0.38$ and the gapless superconductivity
is realized for $\xi>0.38$.
(b) $R_{\rm s}(T)/R_{\rm n}$ vs $T/T_{\rm c}$ in a logarithmic scale.
(c) Superfluid density $\rho_{\rm s}/\rho_{\rm n}$ vs $T/T_{\rm c}$.
}
\end{figure}
%%%%%%%%%%%%%%%%%%%%%%%%%%%%%%%%%%%%%%%%%%%%%%%%%%%%%

\subsection{Nodeless $d$-wave gap}

Next, we move onto the nodeless $d$-wave gap function.
We solve the self-consistent equations Eqs.~(\ref{sol0}), (\ref{sol1}), (\ref{sol2}),
and (\ref{sol3}) for the nodeless $d$-wave gap eq.~(\ref{dgap}).
Owing to the property of the sign change among the different
pocketlike Fermi-surface curves, $g_1=0$ is the trivial solution.
Thus, we solve the self-consistent equation for ${\tilde \omega}$ as
\begin{equation}
{\tilde \omega}=\omega-\xi\frac{\sqrt{1-{\tilde \omega}^2}}{{\tilde \omega}}.
\end{equation}
The DOS $N(\omega)$ is given by
\begin{equation}
N(\omega)=-{\rm Im} g_0 = {\rm Im} \frac{{\tilde \omega}}{\sqrt{1-{\tilde \omega}^2}}.
\end{equation}
Note that in general, ${\tilde \omega}$ becomes a complex number.

In Fig.~8(a), we show $N(\omega)$ for the nodeless $d$-wave gap
for various values of $\xi$.
For the case without impurities ($\xi=0$),
the DOS is zero for $0<\omega<1$, while for $\omega>1$,
we obtain
\begin{equation}
N(\omega)= \frac{\omega}{\sqrt{\omega^2-1}},
\end{equation}
which is the same as that for the simple $s$-wave case.
For $\xi > 0$, we find the finite DOS due to
resonant scattering near the Fermi level.
It is well known that the finite DOS appears at the Fermi level
when we include the impurity scattering
in the self-consistent $T$-matrix approximation in the case of the nodal gap.
The present calculations indicate that even for the nodeless $d$-wave gap,
a finite DOS appears at the Fermi level.
The important point is the phase change of the gap on the Fermi-surface curve,
not the existence of the node.

When we increase the impurity concentration, the value of $\xi$ increases.
To clarify the change in the gap due to the impurity,
we evaluate the values of the gap edges,
$\omega_1$ and $\omega_2$.
First, we transform the self-consistent equation concerning ${\tilde \omega}$
to the quartic equation in terms of $g_0$.
Then, we identify the condition to obtain real solutions of $g_0$,
corresponding to the region of $N(\omega)=0$ with the gap edges.
Note that $\omega_1=0$ and $\omega_2=1$ for $\xi=0$.
As observed in the inset of Fig.~8(a),
the gap size defined by $\omega_2-\omega_1$ becomes zero at $\xi=0.38$.
It is emphasized that even if the gap disappears in the DOS,
the superconductivity is not perfectly destroyed,
leading to the gapless superconductivity.
The impurity-induced gapless superconductivity has been found
in the $s$-wave gap with paramagnetic impurities.\cite{Maki}
In this sense, the present result provides another example of
gapless superconductivity.

Next, we show the calculated results on nuclear magnetic relaxation rate
and superfluid density.
In Fig.~8(b), we show the temperature dependence of $R_{\rm s}/R_{\rm n}$.
In the case of $\zeta=0$, we introduce a cut-off by hand to avoid the divergence
in the DOS at $\omega=1$, since it brings about
the divergence in $R_{\rm s}/R_{\rm n}$ just below $T=T_{\rm c}$.
In the experiments for actual superconducting materials,
it appears as the coherence peak just below $T=T_{\rm c}$.
We also observe the exponential decay in $R_{\rm s}/R_{\rm n}$
at low temperatures.
These properties are characteristic of the $s$-wave superconductivity.

Now, we consider the impurity effect on $R_{\rm s}/R_{\rm n}$.
For $\zeta>0$, owing to the appearance of the finite DOS
near the Fermi level, we observe an almost constant behavior
in $R_{\rm s}/R_{\rm n}$ at low temperatures,
suggesting the revival of the Korringa law at low temperatures.
This is one of the common characteristic issues for the nonmagnetic impurity
effect in unconventional superconductivity.
$R_{\rm s}/R_{\rm n}$ totally increases with the increase of $\zeta$.
Here, we note again that there exists an islandlike finite DOS near the Fermi level.
When $\zeta$ is smaller than the value at which the gap disappears,
the gap exists in the region of $0<\omega<1$.
Thus, $R_{\rm s}/R_{\rm n}$ slightly decreases
with the increase in $T$ in the low-temperature region,
while for $T$ larger than the lower gap edge,
$R_{\rm s}/R_{\rm n}$ increases.
In the dirty nodeless $d$-wave superconductors,
we expect such nonmonotonic temperature dependence
in $R_{\rm s}/R_{\rm n}$, although this behavior is not so significant.

Now, we turn our attention to the superfluid density in Fig.~8(c).
For $\zeta=0$, we find that the temperature dependence
of $\rho_{\rm s}/\rho_{\rm n}$ is similar to that of
the simple $s$-wave superconductor.
For $T<0.3~T_{\rm c}$, $\rho_{\rm s}/\rho_{\rm n}$ is almost unity.
When $T$ is further increased, $\rho_{\rm s}/\rho_{\rm n}$ begins to
decrease and eventually becomes zero at $T=T_{\rm c}$.
For $\zeta > 0$, owing to the appearance of the finite constant
DOS near the Fermi level,
the magnitude of $\rho_{\rm s}/\rho_{\rm n}$ decreases,
but we still observe the almost constant value of $\rho_{\rm s}/\rho_{\rm n}$
at low temperatures.
In this sense, for both clean and dirty cases, the $s$-wave-like
temperature dependence of $\rho_{\rm s}/\rho_{\rm n}$
is expected in the nodeless $d$-wave superconductivity,
when we normalize $\rho_{\rm s}$ by its value at $T=0$.

For small $\zeta$, owing to the existence of the gap in the region of
$0<\omega<1$,
we also find the nonmonotonic temperature dependence of
$\rho_{\rm s}/\rho_{\rm n}$.
The origin is considered to be the same as that in $R_{\rm s}/R_{\rm n}$.
In principle, it is possible to distinguish $\rho_{\rm s}/\rho_{\rm n}$
between clean and dirty cases by the temperature dependence,
but it seems to be difficult to detect such subtle changes
at low temperatures in actual experiments.
It is more realistic to check whether the almost constant part appears or not
in $R_{\rm s}/R_{\rm n}$ at sufficiently low temperatures.

\section{Discussion and Summary}

In this paper, in order to obtain some hints to clarify the node structure
in BiS$_2$-based layered superconductors,
we have investigated the nonmagnetic impurity effect
within the self-consistent $T$-matrix approximation.
We have performed the calculations of the DOS,
nuclear magnetic relaxation rate, and superfluid density
in the superconducting state.

We have assumed two cases, namely, the extended $s$- and $d$-wave gap functions.
Note that in the present material, we have considered the disconnected pocketlike
Fermi-surface curves at around $\mib{k}=(\pm \pi, 0)$ and $(0,\pm \pi)$.
Owing to this structure, there exists no node for the $d$-wave gap,
since the nodal lines along the lines of $k_x=\pm k_y$ do not cross
the Fermi-surface curves.
However, a sign change occurs between the gaps on the pocketlike
Fermi-surface curves.
Thus, nonmagnetic impurities affect seriously the nodeless $d$-wave gap.

On the other hand, for the extended $s$-wave gap,
nodes appear on the Fermi-surface curves.
In particular, for the case without the isotropic part, the nonmagnetic
impurity effect is essentially the same as that in the nodal $d$-wave superconductors.
In this work, we have investigated the impurity effects
for all the cases with the isotropic part.

We have considered two distinct experimental results
concerning the gap nodes on the Fermi-surface curves.
When we consider the extended $s$-wave gap with $p=0$,
the node structure is consistent with the ARPES results.
Moreover, if BiS$_2$-based layered superconductors are assumed to be
dirty owing to the existence of randomness and/or lattice mismatch,
even if nonmagnetic impurities are not explicitly doped,
the resonant scattering effect induces the finite DOS
near the Fermi level, leading to the $s$-wave-like behavior in
physical quantities at low temperatures.
For instance, the temperature dependence of $\rho_s/\rho_n$
in the low-temperature region is similar to that of the $s$-wave case,
if we normalize $\rho_s/\rho_n$ by its value at $T=0$.
In order to confirm the above idea of the dirty
nodal extended $s$-wave superconductivity,
we propose the performance of the measurement of $T_1^{-1}$
in BiS$_2$-based layered superconductors.
The detection of the Korringa-like behavior at low temperatures
will be a test of  the dirty nodal extended $s$-wave superconductivity.

Now, let us consider another possibility that the ARPES experiments
do not indicate the existence of the nodes of the gap.
If we accept this possibility, readers may be inclined to
think that the simple $s$-wave superconductivity is the unique solution.
However, as shown in this paper,
a possibility of the nodeless $d$-wave superconductivity cannot be excluded
in the BiS$_2$-based layered superconductors,
since the temperature dependence of physical quantities
in the nodeless $d$-wave superconductor seems to be
the same as that in the conventional $s$-wave one
in the case of the disconnected Fermi-surface structure
for both clean and dirty cases.

For the confirmation of the nodeless $d$-wave superconductivity,
it is necessary to perform an experiment in which the sign change
between the gaps on the Fermi-surface curves is detected.
Thus, the non-magnetic impurity effect can be the key experiment.
The finite DOS appears due to a small amount of
the nonmagnetic impurity or randomness in the nodeless
$d$-wave superconductor.
Namely, the detection of the Korringa-like behavior at low temperatures
in the dirty case will be the test of the nodeless $d$-wave superconductivity,
provided that we ignore the existence of the nodes on the gap.
Note that the nonmonotonic temperature dependence in
$\rho_s/\rho_n$ and $R_s/R_n$ is interesting from a theoretical
viewpoint, but in actual experiments, it seems difficult to
detect such subtle changes in the temperature dependence
of physical quantities.

Concerning the mechanism of superconductivity in BiS$_2$-based layered materials,
it has not been confirmed yet, but the gap symmetry has been vigorously investigated.
It was suggested by some theoretical researchers that the gap function in
BiS$_2$-based layered superconductors could be
explained by extended $s$-wave, $d$-wave, triplet or
other unconventional pairing scenarios.\cite{Usui,Suzuki,Aga,Martins,Zhou,Yang,Liang,Wu}
On the other hand, a possibility of $s$-wave pairing due to electron-phonon
interaction has been discussed.\cite{Yildirim,Wan,Li,Morice}
The appearance of superconductivity has been discussed from various
theoretical viewpoints on the basis of this two-orbital Hubbard model 
\cite{Usui,Suzuki,Aga,Martins,Zhou,Yang,Liang,Wu}.
The effects of electron-phonon interaction \cite{Yildirim,Wan,Li,Morice}
and spin-orbit coupling \cite{Gao} have also been discussed.
The relationship between the characteristic change of the
Fermi-surface topology and the symmetry of the superconducting gap function
has been pointed out in previous works.\cite{Usui,Suzuki,Aga,Martins,Zhou,Yang,Liang}
It has been discussed that the impurity effect can be a probe of
the pairing symmetry in BiS$_2$-based layered superconductors.\cite{Lin}

The gap anisotropy in BiS$_2$-based layered materials
has been discussed in the context of multiorbital superconductivity.
\cite{Griffith,Suzuki2}
In this paper, we have not included orbital degrees of freedom
in the impurity scattering, but the gap state is considered to be
affected, more or less,
by the nonmagnetic impurity scattering even in multiorbital superconductors,
when the gap has nodes on the Fermi-surface curve or
the sign is reversed between the gaps on the different Fermi-surface curves.
Thus, we believe that the impurity effect is also useful
for the determination of the gap symmetry of multiorbital superconductors,
although it is necessary to develop carefully the discussion
on the nonmagnetic impurity scattering in multi-orbital superconductors
from a quantitative viewpoint.

Finally, we comment on the nodeless $d$-wave superconductivity in other systems.
The nodeless $d$-wave gap was discussed in a quasi-one-dimensinoal
organic superconductor.\cite{Shimahara}
A possible high-$T_{\rm c}$ mechanism due to spin fluctuations was proposed
in a system with a Fermi-surface pocket.\cite{Kuroki}
In electron-doped high-$T_{\rm c}$ cuprates, nodeless $d$-wave
superconductivity was also examined.\cite{Yuan,Das}
Quite recently, the nodeless $d$-wave gap has been considered
in monolayer FeSe.\cite{Agterberg}
Here, we do not mention the superconducting mechanism
of iron-based superconductors, since it is beyond the scope of this paper,
but we refer to two papers, Refs.~\citen{Senga} and \citen{Bang},
in which the impurity effects in the $s_{\pm}$-wave superconducting
state with a multiband structure were discussed.

In summary, to reconcile the existence of the gap nodes
on the Fermi-surface curves and
the $s$-wave-like temperature dependence of physical quantities
in BiS$_2$-based layered superconductors,
we have proposed the dirty nodal extended $s$-wave gap
without an isotropic part.
Note that in this scenario, it is necessary to assume that
the sample becomes dirty owing to the existence of randomness
and/or lattice mismatch,
even if nonmagnetic impurities are not explicitly doped.
Provided that the existence of the node can be ignored,
we have suggested a nontrivial possibility of
nodeless $d$-wave superconductivity for both clean and dirty cases,
in addition to the conventional $s$-wave superconductivity.
In both scenarios, a key experiment will be the measurement of $T_1^{-1}$
in BiS$_2$-based layered superconductors.

\section*{Acknowledgments}

The authors thank Y. Aoki, K. Kubo, T. Matsuda, Y. Mizuguchi, K. Hattori,
and R. Higashinaka for discussions on BiS$_2$-based layered materials.
This work has been supported by KAKENHI (16H04017).
The computation in this work was partly carried out using the facilities of the
Supercomputer Center of the Institute for Solid State Physics,
University of Tokyo.

\end{document}